\documentclass[sigconf]{acmart}

\usepackage{url}
\usepackage{balance}  
\usepackage{comment}
\usepackage{listings}
\usepackage{graphicx}
\usepackage{xcolor}
\usepackage{subcaption}
\usepackage{algorithm}
\usepackage{algorithmic}

\DeclareMathOperator*{\argmax}{arg\,max} 

\AtBeginDocument{%
  \providecommand\BibTeX{{%
    \normalfont B\kern-0.5em{\scshape i\kern-0.25em b}\kern-0.8em\TeX}}}

\setcopyright{acmcopyright}
\copyrightyear{2018}
\acmYear{2018}
\acmDOI{10.1145/1122445.1122456}

\acmConference[Woodstock '18]{Woodstock '18: ACM Symposium on Neural
  Gaze Detection}{June 03--05, 2018}{Woodstock, NY}
\acmBooktitle{Woodstock '18: ACM Symposium on Neural Gaze Detection,
  June 03--05, 2018, Woodstock, NY}
\acmPrice{15.00}
\acmISBN{978-1-4503-XXXX-X/18/06}

\begin{document}

\title{DrugDBEmbed : Semantic Queries on Relational Database using Supervised Column Encodings}

\author{Bortik Bandyopadhyay, Pranav Maneriker, Vedang Patel, Saumya Yashmohini Sahai, Ping Zhang, Srinivasan Parthasarathy}
\affiliation{
  \institution{The Ohio State University}
  \city{Columbus, Ohio}}
\email{{bandyopadhyay.14, maneriker.1, patel.3140, sahai.17, zhang.10631}@osu.edu, srini@cse.ohio-state.edu}

\begin{abstract}
Traditional relational databases contain a lot of latent semantic information that have largely remained untapped due to the difficulty involved in automatically extracting such information.
Recent works have proposed unsupervised machine learning approaches to extract such hidden information by textifying the database columns and then projecting the text tokens onto a fixed dimensional semantic vector space.
However, in certain databases, task-specific class labels may be available, which unsupervised approaches are unable to lever in a principled manner.
Also, when embeddings are generated at individual token level, then column encoding of multi-token text column has to be computed by taking the average of the vectors of the tokens present in that column for any given row.
Such averaging approach may not produce the best semantic vector representation of the multi-token text column, as observed while encoding paragraphs or documents in natural language processing domain.
With these shortcomings in mind, we propose a supervised machine learning approach using a Bi-LSTM based sequence encoder to directly generate column encodings for multi-token text columns of the DrugBank database, which contains gold standard drug-drug interaction (DDI) labels.
Our text data driven encoding approach achieves very high Accuracy on the supervised DDI prediction task for some columns and we use those supervised column encodings to simulate and evaluate the Analogy SQL queries on relational data to demonstrate the efficacy of our technique.\looseness=-1
\end{abstract}


\maketitle

\section{Introduction}

Traditional relational database systems support a wide variety of well-defined structured and unstructured data types like numeric, categorical, unstructured text, images etc.~\cite{Bordawekar2016, Bordawekar2017}.
The relational data can be accessed by SQL, which only allows for syntactic matching or range-based queries~\cite{Bordawekar2017} through select, project and join operations.
However, as pointed out by Bordawekar and Shmueli~\cite{Bordawekar2016}, there are a lot of latent semantic information present inside a relational database that cannot be directly utilized through SQL queries on the original data. 
The latent semantic information could sometimes be at individual token level, while in other cases, a sequence of text tokens can jointly convey that information~\cite{Bordawekar2016}.
While there are dictionary-based text-extenders~\cite{Cutlip2003} or ontology-based support systems ~\cite{Lim2013semantic} to partially enable semantic queries, such systems cannot directly extract the latent semantic information from the database~\cite{Bordawekar2016}.
Bordawekar and Shmueli~\cite{Bordawekar2016} propose a textification based database token embedding generation approach to automatically extract latent semantic information from a relational database, by projecting the \textit{textified database tokens} to a low dimensional vector space using an unsupervised token embedding generation technique, adapted from the popular word2vec algorithm~\cite{Mikolov2013}. Cosine similarity between such vector representations of database tokens can be easily computed via custom user defined functions as part of SQL queries~\cite{Bordawekar2017}.
Additionally, those vector representations can be used to run a wide variety of semantic queries~\cite{Bordawekar2017} like approximate nearest neighbor query, Analogy SQL query etc.\looseness=-1

The unsupervised token embedding generation approach proposed by Bordawekar et al.~\cite{Bordawekar2017} is based on the distributional hypothesis of words in text corpora~\cite{Mikolov2013}, which makes them generic and well suited for exploratory data analytics.
In contrast, we believe a supervised embedding generation approach maybe more suited to task-specific scenarios, where ground-truth labels are available. 
Our assumption of databases containing task-specific supervision information is grounded on practical examples.
Take for example, a researcher who is working on a novel drug discovery problem in a pharmaceutical company. She has access to a highly curated proprietary relational database that contains information of prescription as well as under-evaluation drugs, along with gold standard drug-drug interaction information~\footnote{Drug-Drug interaction implies adverse side-effects when two prescription drugs are taken together by a patient. Detecting possible drug-drug interaction is an important task, since it can save a lot of human lives and control annual health care costs~\cite{Giacomini2007good, Zheng2017attention}\looseness=-1}.
Although she has little or no expertise of SQL query, the researcher may be interested in retrieving from such a proprietary relational database, the list of all drugs that interact with a query drug A in the same way as B interacts with A. 
Such an end-user query can be executed on the database using drug embeddings, through an approximation-based Analogy SQL query~\cite{Bordawekar2017}.
However, it would be preferred that the drug embeddings are \textit{task-specific} i.e., they are generated to capture the gold standard drug-drug interaction information, instead of being task-agnostic.\looseness=-1

To this end, we propose a Bi-LSTM based supervised column encoding generation approach for multi-token text columns of a relational database using techniques from the natural language processing domain~\cite{Kim14cnn, LiuQH2016, Yang2016, Zhou2016text} and utilize such embeddings to solve Analogy SQL queries~\cite{Bordawekar2017} on the relational database.
For exposition simplicity, we describe our approach using the Drug database example, where the ground truth drug-drug interaction labels are present in a special table (DDI) in addition to the table containing drug information (DI). 
Each row of the DDI table contains an interaction type (class label) for a pair of rows (i.e., the corresponding drugs) in the DI table.
We utilize the drug pair interaction label to generate column encodings for the text-based column of the corresponding drug rows in the DI table, by training a classification model for the DDI prediction task.
Thus, we generate \textit{database column encodings based on relationship between a pair of rows} using ground truth labels.
We train our classification model from scratch, which gives the embedding of individual tokens as a by-product of our approach.
Thus the column and token encodings, generated by our proposed Bi-LSTM model, are fully task-specific interaction relationship-based encodings, as opposed to the generic word distribution based encodings of ~\cite{Bordawekar2017}.\looseness=-1

To the best of our knowledge, our approach is the first to \textit{explicitly utilize gold-standard row-pair relationship label driven supervised column encodings} for semantic queries on relational databases using the drug drug interaction use-case.
Prior works like~\cite{Bordawekar2016, Bordawekar2017, Bordawekar2017using, Bordawekar2019} have focused on generating task-agnostic unsupervised embeddings for each individual token of a sequence of text tokens within a database row. 
In such a setting, one way to compute column encoding for a multi-token column is to compute the average of the vectors of the text tokens present in that column~\cite{Bordawekar2017}, which may not always be the best semantic vector representation for that text column~\cite{Le2014distributed}.
In contrast, we focus on directly generating \textit{task-specific supervised column encodings} i.e., a vector representation for a sequence of text tokens present within the column boundary, similar to paragraph~\cite{Le2014distributed} and document~\cite{Dai2015document} encoding generation approaches in NLP, albeit in a supervised learning setting.
We get embeddings for individual tokens as a by-product of our approach. 
Since we demonstrate our approach using DDI scenario, it is important to point out that there are several works on the supervised DDI prediction task~\cite{Zhang2015label, Fokoue2016predicting, Ryu2018deep,Kastrin2018predicting,Zitnik2018modeling, Deng2020}, but most of them focus on the chemical structure of the drugs. 
For the works~\cite{Zhang2015label, Fokoue2016predicting, Kastrin2018predicting} that use text based information, there is a costly data pre-processing and feature extraction phase.
In contrast, we focus on only categorical and text based attributes and our DDI prediction approach can take the text token sequence as input after a very light-weight text pre-processing step, which significantly reduces the overhead of the feature generation step for the classification task.
Our proposed Bi-LSTM model achieves very competitive performance over standard BOW baselines for the classification task.
We demonstrate the efficacy of our supervised column encodings for the Analogy SQL task, by proposing an intuitive DDI pair based simulation and evaluation strategy, under two different real-world settings.
\looseness=-1

\section{Related Work}

\textbf{\noindent{Language Embedding:}} In the NLP domain, language embedding generation, for obtaining a vector representation of words in a language, has been extensively studied. A few such techniques include neural network based learning~\cite{bengio2003neural}, log-linear classifiers ~\cite{Mikolov2017}, function optimization of objective matrix~\cite{Pennington2014}, matrix  factorization techniques~\cite{levy2014neural} etc. The unsupervised database token embedding generation technique used by Bordawekar et al.~\cite{Bordawekar2017} is an adaptation of one of the most popular word embedding algorithm called word2Vec~\cite{Mikolov2013,DBLPMikolovLS13}. The vectors produced by word2Vec~\cite{Mikolov2013,DBLPMikolovLS13} can be utilized for many semantic reasoning tasks like analogy~\cite{Levy2014} with good performance. Additional details about word2Vec can be obtained here~\cite{mnih2013learning,DBLPGoldbergL14,Levy2014} while its modifications for databases are detailed here~\cite{Bordawekar2017}.\looseness=-1

Paragraphs~\cite{Le2014distributed} and documents~\cite{Dai2015document} can also be summarized using vector representation for various downstream applications. 
For text classification tasks, BOW models have been used as baselines~\cite{Wang2012baselines}. 
However, there are many deep learning techniques like the CNN~\cite{Kim14cnn},  RNN~\cite{LiuQH2016}, LSTM~\cite{Yang2016} and Bi-LSTM~\cite{Zhou2016text} based models, that can be utilized to generate a fixed low dimensional representation of the input text and use it for the classification task. 
Yin et al.~\cite{Wenpeng2017} has conducted a detailed comparative study of CNN and RNN for such tasks. 
Note that, our goal is to learn vector representations for text token sequence in such a way that we learn the latent relationship between a pair of text expressed by the class label (i.e., DDI type), similar to relation learning tasks~\cite{Lin2016neural, Rossiello2019, Kuang2019improving}.\looseness=-1

\textbf{\noindent Database Embedding generation and utilization: } To address the problem of resolving schematic differences between objects in multiple databases, Kashyap and Sheth~\cite{Kashyap1996} define the semantic proximity between two related objects, by associating the mapping of those objects with the comparison context. The DLDB~\cite{Pan2004dldb} system has been designed to allow semantic web queries on relational databases using the FaCT description logic reasoner. Semantic queries through dictionary-based text-extenders (e.g.: DB2 Text Extender~\cite{Cutlip2003}) or ontology-based support systems~\cite{Lim2013semantic} has been in practice. Yu et al.~\cite{Yu2007} define keyword relationship summaries and utilize those summaries to develop novel ranking methods to select the most relevant database for a given keyword query. However, none of these works use database token embeddings for executing semantic reasoning queries~\cite{Bordawekar2017} using SQL UDF's.\looseness=-1

Bordawekar and Shmueli~\cite{Bordawekar2016} introduce unsupervised machine learning technique to generate database token embeddings that can capture the latent semantic information between database entities. It is a two step process. The first step involves converting heterogeneous data types of the original database (like numeric, text, images etc.) into a consistent text-based representation~\cite{Bordawekar2016, Bordawekar2017}. The database is converted into a text corpus where each line of the text corpus represents a textified row of the original relational database. The authors adapt the very popular word2vec algorithm~\cite{Mikolov2013}, an unsupervised approach, to generate embeddings of the tokens of the textified database corpus. The cosine similarity between any pair of database token (primary key) vectors is the approximate semantic similarity between those tokens (the rows).  Furthermore, the authors demonstrate~\cite{neves2018demo} how custom user defined functions can be incorporated as part of SQL queries to realize enhanced semantic querying capabilities~\cite{Bordawekar2016, Bordawekar2017} like analogy queries, semantic clustering etc. which can be executed by leveraging this low dimensional representation of database tokens. A sample case-study~\cite{Bordawekar2017} demonstrates the usefulness of the novel insights that such semantic queries can uncover from relational databases using simple SQL syntax. The semantic vectors are primarily based on the information present within the database, with the added option of utilizing information from external corpus like Wikipedia~\cite{Bordawekar2017} for semantic querying purposes. 
One advantage of the unsupervised approach is that the embeddings can be incrementally trained~\cite{Bordawekar2017} in case of any database or content updates, which makes it very flexible in practice.
Additionally, such unsupervised database token embeddings can also be effectively utilized for selectively disclosing database information~\cite{Bordawekar2019}. In contrast, we propose a supervised task-specific database column encoding approach that summarizes the information present in database columns wrt the specific task at hand, while also generating embeddings for individual database tokens. \looseness=-1

The sub-problem of approximate nearest neighbor search using token embeddings in a database have been further optimized~\cite{Gunther2019} in the Freddy~\cite{Freddy2018} framework, which is a free open sourced implementation of semantic querying capabilities on top of Postgresql database. 
Cappuzzo et al.~\cite{Cappuzzo2019} first construct a graph based representation of the relational data and then lever unsupervised graph embedding approach to generate database token embeddings, which outperforms over several baseline for the data integration task. The idea of generating an embedding based representation of each row of a database has been also utilized for the missing value imputation task~\cite{Biessmann2018} quite successfully. 
Srinivas et al.~\cite{Srinivas2018} propose a `siamese  triplet' network that assigns a small distance to the two related surface forms of the same entity, which can be utilized very effectively for  improving performance of data merging tasks.
Termite~\cite{Fernandez2019} is a relational embedding framework for data integration task, where the authors aim to learn a distance metric that can be used to compute the similarity/distance of data coming from text and databases. 
In RETRO~\cite{Gunther2019retro}, database token embeddings are generated by formulating relation retrofitting as a learning problem (similar to~\cite{Faruqui2015retrofitting}), such that relational information present in the database can be combined in a principled way with semantic information present in external text corpus for better token embeddings.
Arora and Bedathur~\cite{Arora2020embeddings} propose an LSTM based  model for capturing \textit{inter-row relationships} in relational databases and utilize such embeddings for similarity queries and data cell completion tasks.
Their work clearly demonstrates the advantage of using an LSTM encoder for capturing the sequence based information (temporal information in their setting) over traditional approaches.\looseness=-1

\noindent\textbf{Drug-Drug Interaction Prediction : }
Two drugs when consumed together may cause unexpected adverse side effects. 
Thus, detecting possible adverse drug-drug interaction is a very important task, which can save a lot of human lives as well as, control annual health care costs~\cite{Giacomini2007good, Zheng2017attention}. 
There has been a significant amount of research on the DDI prediction task, but most of those works often involve costly pre-processing and feature extraction step.
Ryu et al. develop the DeepDDI framework~\cite{Ryu2018deep}, which uses the chemical structures of the input drug pair in simplified molecular-input line-entry system (SMILES) format to construct SSP’s for each drug through a multi-step feature construction process and use those SSP vectors as input to train a multilayered perceptron that learns to predict multiple DDI types for the given drug pair. 
Deng et al. develop the DDIMDL framework~\cite{Deng2020} that uses multiple different drug information like chemical structures, targets, enzymes and pathways, to construct corresponding similarity matrices after an elaborate feature extraction and similarity computation step.
The similarity matrices are then used to train a modular DNN network which combines individual sub-model's prediction to generate the final DDI prediction.
In contrast, we use only text data as input, and lever a BiLSTM to do automatic input feature extraction for the one-out-of-many DDI class prediction problem.\looseness=-1

Zhang et al.~\cite{Zhang2015label} propose a label propagation framework for DDI prediction by integrating heterogeneous information (like chemical structure, side effects from package inserts of drugs etc.) from diverse sources.  
Fokoue et al.~\cite{Fokoue2016predicting} design Tiresias, where semantic similarity scores from multiple diverse information sources are computed for drug pairs, and then these scores are used as features to train a logistic regression classifier to predict DDI between the drug pairs. 
Kastrin at al.~\cite{Kastrin2018predicting} extract various topological and semantic similarity based features for drug pairs from multiple heterogeneous information sources and utilize machine learning models (like SVM, GBM etc.) to perform DDI prediction.
Ma et al. ~\cite{Ma2018drug} propose an attentive multi-view graph auto-encoders for DDI prediction.
Note that while some of the previous works~\cite{Zhang2015label, Fokoue2016predicting, Kastrin2018predicting} use text data as one of the many information sources, there is a significant pre-processing burden for feature engineering, as different types of pairwise similarity scores need to be computed for each drug pair.\looseness=-1

Zheng et al.~\cite{Zheng2017attention} propose an attention based by BiLSTM model for effective drug-drug interaction extraction (but not DDI prediction) from medical corpus. 
Asada et al.~\cite{Asada2018enhancing} utilize a graph convolution neural network to capture the molecular structure information of drug pairs, which further improves the quality of drug-drug interaction extraction from biomedical text.
Our approach focuses only on the text based drug information, like atc codes, categories, description etc. of each drug, but not the chemical structure, to better understand the true informativeness of the text-only data for the DDI task.
We use a BiLSTM encoder based supervised neural model for column encoding generation (similar to~\cite{Zheng2017attention} without attention), by learning to predict drug-drug interaction from various text-based drug information. 
Thus our feature extraction process for the classification task is automatic due to the BiLSTM encodings, which can be utilized for the execution of semantic queries on relational databases.
Unsupervised embedding based analogy queries have been evaluated on medical data~\cite{Newman2017insights}, but not in the context of relational databases.\looseness=-1

\section{Methodology}
\label{section:methodoloy}

The overall steps are similar to the one proposed by Bordawekar et al.~\cite{Bordawekar2017}, although our textification step is less involved, since we focus only on multi-token text columns, and the encodings we generate are fully supervised.
In Section~\ref{subsection:generatingencoding}, we describe how to generate novel task-specific column encodings and then in Section~\ref{subsection:leveragingencoding} explain how to use such encodings to run the Analogy SQL query~\cite{Bordawekar2017}.\looseness=-1

\subsection{Generating Column Encoding}
\label{subsection:generatingencoding}

We assume that the data has been converted to text form (more details in Section~\ref{subsection:datagathering}), and hence we can employ various text sequence models for encoding. 
Recurrent Neural Network based models, in particular, Long Short-Term Memory (LSTMs)~\cite{hochreiter1997long} have been successfully used to generate text encodings that can generalize across various task~\cite{peters2018deep, howard2018universal}. 
Our drug encoding models are also based on LSTMs. 
First, we describe the procedure we use to generate encodings for each column using LSTM based encodings. 

Consider a sequence of tokens, where each token is mapped to a unique integer~$\in \{0, \dots, N-1\}$, where $N$ denotes the size of the vocabulary consisting of all the tokens in the dictionary. 
By iterating through the entries in the column in the training set, such a dictionary can be constructed.
A special token is also added to encode unseen tokens in the training data. 
Each element in the vocabulary is then mapped to a d-dimensional vector, and each sequence is either truncated or padded with special tokens to make sure that all sequences have the same length. 
Thus for each sequence $s$, we get a corresponding sequence of encoded d-dimensional vectors $[x_{s1}, x_{s2}, \dots x_{sn}]$, where $n$ is the chosen maximum length.
Each encoded sequence $s$ is then passed through a multi-layer LSTM network. 
The corresponding LSTM equations in the forward direction are~\cite{hochreiter1997long,Pytorch2019}:
\begin{align*}
    i_{st}^l &= \sigma(W_{ii}^l x_{st}^{l-1} + b_{ii}^l + W_{hi}^l h_{s(t-1)}^l + b_{hi}^l)\\
    f_{st}^l &= \sigma(W_{if}^l x_{st}^{l-1} + b_{if}^l + W_{hf}^l h_{s(t-1)}^l + b_{hf}^l)\\
    g_{st}^l &= \tanh(W_{ig}^l x_{st}^{l-1} + b_{ig}^l + W_{hg}^l h_{s(t-1)}^l + b_{hg}^l)\\
    o_{st}^l &= \sigma(W_{io}^l x_{st}^{l-1} + b_{io}^l + W_{ho}^l h_{s(t-1)} + b_{io}^l) \\
    c_{st}^l &= f_{st}^l \odot c_{s(t-1)}^l + i_{st}^l \odot g_{st}^l\\
    h_{st}^l &= o_{st} \odot \tanh(c_{st})\\
    x_{st}^l &= h_{st}^l
\end{align*}
where $i$, $f$, $g$, $o$ represent the input, forget, cell, and output gates/dimensions, $t$ represents the time step, $h$ represents the hidden state/dimension, $l$ represents the layer, $\sigma$ is the sigmoid function, and $\odot$ represents Hadamard product. $W$ and $b$ refer to the weights and biases for internal layers. 
As shown in the last equation, output hidden states $h_{st}^l$ are used as the input states for the next layer $x_{st}^l$, with $x_{st}^0$ being the d-dimensional encoding described previously.
A reverse direction LSTM ($t+1$ replaces $t-1$ in all equations) output is computed in parallel with the output of this LSTM. 
For an $L$ layer network, the final hidden states ($h_{sN}^{L(f)}$ for forward and $h_{0N}^{L(b)}$ for the backward) are concatenated and used as the representation for the entire sequence.
We label the output ${e_s = h_{sN}^{L(f)} \oplus h_{0N}^{L(b)}}$, as the embedding for sequence $s$, where $\oplus$ represents the concatenation operation.
To tune these embeddings for predicting drug drug interactions, we use ideas from relation learning ~\cite{wang2014knowledge}. For each pair of drugs encodings $e_{i}, e_{j}$ in the training data, we construct a pair encoding as 
\begin{align*}
    p_{ij} = (|e_i - e_j|) \oplus (e_i \odot e_j)
\end{align*}
\begin{figure*}
    \centering
    \includegraphics[width=\linewidth]{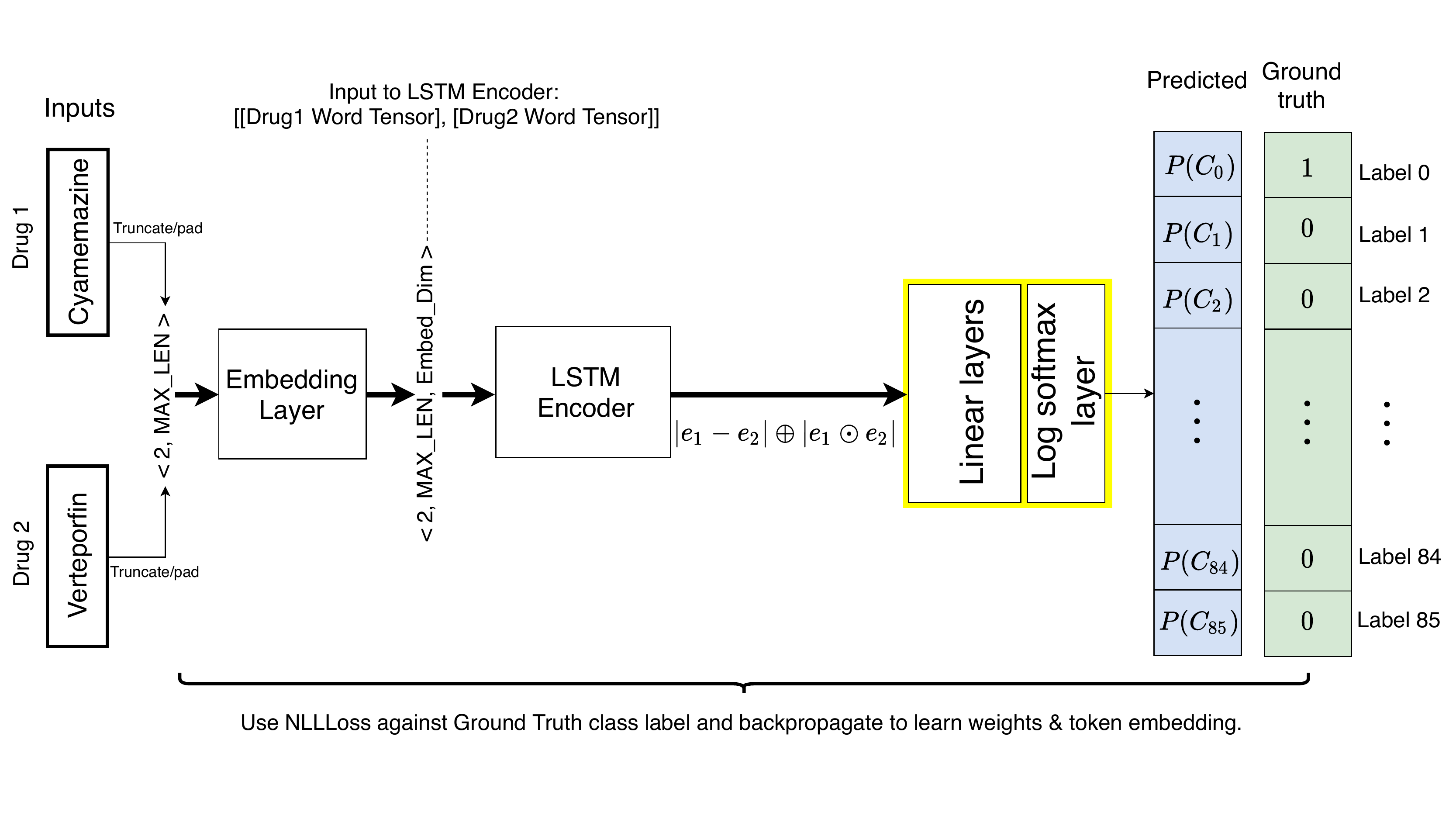}
    \vspace{-1cm}
    \caption{Overview of the proposed DNN model for column encoding generation via learning the DDI prediction task.}
    \label{fig:lstm_encoding}
\end{figure*}
Prior work~\cite{wang2014knowledge} has shown that using the difference and product can help capture the relationship between pairs of encodings. 
Finally, we pass the pair encoding through a single linear layer to get the logit scores corresponding to each interaction label.
Figure~\ref{fig:lstm_encoding} demonstrates the complete series of steps used to get the scores for each interaction class for a single drug pair. 
These scores are then converted into log softmax values after normalization and then used to compute a negative log likelihood based loss for predicting the correct interaction class.
We use a single hidden layer to avoid offloading any complexities of interactions in the final layers - the encodings themselves contain the information corresponding to the interactions.\looseness=-1

\subsection{Leveraging Column Encoding}
\label{subsection:leveragingencoding}

The Deep Learning based classification model developed above can be used to generate task-specific (i.e., DDI interaction prediction based) d-dimensional column encodings for each of the text columns of each drug, by concatenating the final bi-directional hidden states of the trained Bi-LSTM model.
Then the column-specific encodings are uploaded in separate tables (called encoding tables) in the database, where the primary key is the drug id and the column contains the d-dimensional vector~\cite{Bordawekar2017, Freddy2018}.
Since the encodings are column specific, hence Analogy SQL queries~\cite{Bordawekar2017} on drugs can now be executed by leveraging the relevant column encodings of the query drugs.
We now present a brief overview of executing Analogy SQL queries using those encodings, as proposed by Bordawekar et al.~\cite{Bordawekar2017}.\looseness=-1

Given the elements A,B,C,D have been projected as vectors in a common multi-dimensional vector space, Rumelhart and Abrahamson~\cite{Rumelhart1973model} describe analogy query ($A : B :: C : D$) as the task of finding a vector D, whose distance from the vector of C is closest to the distance between the vectors of A and B.
In our settings, the elements A,B,C,D can be thought of as unique drug names (or drug identifiers)  which have been projected into a common vector space based on the text data of the specific column for individual drugs (as described in Section~\ref{subsection:generatingencoding}).
While many standard techniques exist in the NLP literature for Analogy query computation using word embeddings, we use the cosine similarity ($Cosine$) based 3COSMUL~\cite{Levy2014} strategy, as it has shown consistently good performance in both NLP~\cite{Levy2014} and Database~\cite{Bordawekar2017} settings for analogy computation task.
For completeness, we present the formula to compute 3COSMUL~\cite{Levy2014,Bordawekar2017} for a given analogy query $A : B :: C : D$, and the corresponding vectors $V_A$, $V_B$, $V_C$, $V_D$ below as: 
\begin{equation}\label{eq:3cosmul}
    \argmax_{D \in Drugs} \frac{C(V_D, V_C) * C(V_D, V_B)}{C(V_D, V_A) + \epsilon}
\end{equation}
where, $C(V_1, V_2) = (Cosine(V_1, V_2) + 1.0)/2.0$ to make the final analogy score as positive for maximization purpose and $\epsilon = 0.001$ to avoid division by zero error~\cite{Levy2014}.
The method 3COSMUL is implemented as part of SQL query in the Freddy framework~\cite{Freddy2018} to realize the column specific analogy functionality utilizing the respective column embedding tables.
A sample analogy query (`DB08897' : `DB11315' :: `DB08897' : ?) with real drugs using one of the column's encodings (\textit{ColEncTable}) is shown below.

\begin{lstlisting}[language=SQL, showspaces=false, frame = single, basicstyle=\tiny]
SELECT DISTINCT T.drugbank_id as DrugId, 
   (C(v4.vector, v3.vector) * C(v4.vector, v2.vector))/
   (C(v4.vector, v1.vector) + 0.001) AS Score
FROM DrugBankFullClean AS T
INNER JOIN ColEncTable AS v1 ON v1.drugbank_id = 'DB08897'
INNER JOIN ColEncTable AS v2 ON v2.drugbank_id = 'DB11315'
INNER JOIN ColEncTable AS v3 ON v3.drugbank_id = 'DB08897'
INNER JOIN ColEncTable AS v4 ON v4.drugbank_id = T.drugbank_id
WHERE T.drugbank_id NOT IN ('DB08897', 'DB11315', 'DB08897')
AND
	(C(v4.vector, v3.vector) * C(v4.vector, v2.vector))/
	(C(v4.vector, v1.vector) + 0.001)  >= 0.25
ORDER BY (C(v4.vector, v3.vector) * C(v4.vector, v2.vector))/
    (C(v4.vector, v1.vector) + 0.001) DESC
FETCH FIRST 10 ROWS ONLY;
\end{lstlisting}

In the above query, note that A and C are same and only top-10 most relevant drugs are retrieved as the answer drug for the given analogy query, such that the answer drugs drugD presumably have the same interaction type (i.e., DDI label) with drug drugC (`DB08897') as the interaction type between the drug pair drugA (`DB08897') and drugB (`DB11315').
We describe the simulation and evaluation strategy for such queries developed by us in Section~\ref{subsection:analogysimulation} and Section~\ref{subsection:evalmetrics} respectively.

\section{Experiment Setup}
\label{section:experiments}

We implement all components of our framework in Python, and use Freddy~\cite{Freddy2018} as the back-end database.
The BOW based models and the DNN model have been implemented using \texttt{scikit-learn}~\cite{scikit-learn} and PyTorch~\cite{Pytorch2019} respectively. 
Experiments are run on machines hosted at Ohio Supercomputer Center~\cite{Pitzer2018}.\looseness=-1

\subsection{Data Gathering}
\label{subsection:datagathering}

\subsubsection{Extracting the Data}
\label{subsubsection:dataextraction}

We use the publicly available parser~\footnote{https://github.com/dhimmel/drugbank} to extract the raw drug information downloaded from DrugBank\footnote{\url{https://www.drugbank.ca/}}.
We focus on 6 text based attributes viz: ATC codes, Categories Description, Merged Class information, Protein Binding and Target Action.
We concatenate various class based information of each drug into a single categorical attribute that we call as Merged Class information in the rest of the paper.
Our approach will work for any columns that are of categorical or text type or can be easily textified by various methods proposed in~\cite{Bordawekar2017}.
We use the ground-truth drug drug interaction labels from Ryu et al.~\cite{Ryu2018deep} which contains about 192K DDI pairs.
Since our setting is a single class classification problem, we randomly retain only a single DDI interaction for drug pairs that have multiple interaction labels.
We perform very light-weight basic text pre-processing on the raw drug information and retain only those drugs for which there is a valid DDI pair above.
The final data contains textified information of 1705 drugs extracted from the DrugBank corpus, and 191,728 corresponding DDI pairs with total 86 different interaction types (class labels).\looseness=-1

\subsubsection{Basic Text Pre-processing}
\label{subsubsection:textprocessing}

For every input text field, we first convert all characters to lowercase. 
Following this step, we truncate multiple white spaces into a single space for cleaner tokenization, and split the input into tokens based on the space character. 
Further, we remove all words that are present in the NLTK stopwords corpus for English~\cite{Loper02nltk}.
We replace most of the numeric tokens by a special token `numtkn' that indicates the presence of a number (only applicable for certain columns).
For some of the categorical columns, we observe that sometimes each drug may have multiple unique values in a column.
For example, a single drug may have multiple different ATC codes. 
Hence during data extraction phase we separate each unique value of a multi-valued categorical column using a special character (`\#') and later use this separator to individually process each value using custom pre-processing for each column (details for a couple of them presented below).
\begin{itemize}
    \item ATC codes: We first convert all text to lowercase. Then, we split the string by `\#' and extract each ATC code in the string and concatenate them with special indicators denoting which level of the code that they correspond to. Note that each ATC code~\footnote{\url{https://en.wikipedia.org/wiki/Anatomical_Therapeutic_Chemical_Classification_System}} is 7 characters in length and has special ontological information encoded for each level. We want to preserve this ontological structure by pre-pending the level information to each of the code corresponding to that level. Thus for an ATC code of ``B01AE02'' the processed token sequence is (``atcl1\_b atcl2\_01 atcl3\_a atcl4\_e atcl5\_02'').
    \item Categories: We split the input text by `\#', and delete all types of parenthesis characters. We further split each split by comma and slash characters.  
\end{itemize}

We construct the vocabulary for each column by extracting the list of unique tokens in that column for the set of all unique drugs in training and validation partitions.
In addition to the vocabulary consisting of the tokens seen during training (with more than a minimum threshold frequency), we additionally add a special token for unseen (unknown) words, one for padding, and one to mark the end of a sequence of tokens, all of which are common in NLP domain.
Note that our tokenization approach is rather simple and often produces several tokens which may not be found in some of the publicly available pre-trained biomedical corpus embedding, which force us to train our token embeddings from scratch as part of the model training procedure.
This simplicity of tokenization is adapted to make the tokens more human readable, albeit with a possible qualitative performance impact.
We leave the use of more sophisticated sub-word information based tokenization schemes like byte pair encoding~\cite{Gage1994} for future exploration.

\subsubsection{Initializing the Database}
\label{subsubsection:dbinit}

We use the publicly available postgresql based framework called Freddy~\cite{Freddy2018} as the data back-end.
All the textified information of the drugs are loaded in a table called \textit{DrugBankFullClean}, where the DrugBank identifier field of each drug forms the primary key of this table.
The DDI pairs, along with the ground-truth labels, are loaded in the \textit{DDITypeInfoTable} having column names : (drug1, drug2, label).
This table has a composite primary key $<$D1Id, D2Id$>$, where D1Id and D2Id are the DrugBank identifiers for the first (drug1) and the second (drug2) columns of each of the DDI pairs respectively.
Besides these 2 tables , there are separate \textit{ColEncTable} tables, one for each of the 6 columns, prefixed with the column name that store the $<$Drug Id, Column Encodings$>$, where the corresponding column encodings of each drug is generated from the trained DNN model. 
The dimension of the encodings may vary for each column depending on the parameters of the best trained model for the corresponding column.\looseness=-1

\subsection{Data Partitioning}
\label{subsection:datapartitioning}

To train the ML models for the supervised DDI prediction task, the drug pair interaction data consisting of the ground-truth labels, needs to be partitioned into training, validation and testing set.
We create two separate partitioning strategies, as described below, and run all the experiments under both of these settings to evaluate the efficacy of our strategy.\looseness=-1

\subsubsection{DDI Pairwise Partitioning}
\label{subsubsection:ddipartition}

This partitioning approach is a common way to train and evaluate many supervised machine learning models. We use stratified sampling technique, with each class of the 86 class labels as a strata, to split the 191,728 drug pair interaction data into 80\% training (153,346 DDI pairs), 10\% validation (19,171 DDI pairs) and remaining 10\% testing (19,211 DDI pairs) set. Thus, each unique label has similar distribution in all the partitions, which makes model evaluation using the partitions, easier.\looseness=-1

\subsubsection{Drug Held-off Partitioning}
\label{subsubsection:heldoffpartition}

This partitioning approach is rooted in the practical end-user scenario~\cite{Zhang2015label}, where the database already contains several drugs as well as the corresponding ground truth DDI pairs, and then some new drugs (along with the drug's information) get introduced into the database.
Now the task is to predict the DDI interaction of these new drugs using the DDI model and also to execute the Analogy SQL query using the corresponding encodings.
We think this setting is a much harder one to evaluate, compared to the DDI pairwise partitioning scheme, albeit the most practical one.
For this setting, we randomly select x\% of the total unique drugs from our filtered drug set and consider them as held-off drugs.
Then we remove all the ground-truth DDI pairs where \textit{atleast one of the drug belongs to the held-off drug set}, and we designate this as the testing DDI set.
The testing set may not contain drug pairs from all 86 classes and thus the label distribution is skewed.
The \textit{remaining ground truth DDI pairs} are split randomly into Training (90\%) and Validation (10\%) set, ensuring that both of these set contains atleast one drug drug interaction pair from each of the 86 classes.\looseness=-1

\begin{table}[!htb]
\begin{tabular}{|p{4.2cm}|p{1cm}|p{1cm}|p{1cm}|}
\cline{1-4}
& 1\% Held-Off & 2\% Held-Off & 3\% Held-Off\\
\cline{1-4}
\#Unique Total Drugs & 1705 & 1705 & 1705\\
\hline
\#Unique held-off Test Drugs & 17 & 34 & 51\\
\hline
\#Unique Train + Val Drugs & 1688 & 1671 & 1654\\
\hline
\hline
\hline
\#Training Drug Pairs & 169,565 & 165,446 & 160,951\\
\hline
\#Unique Labels in Training & 86/86 & 86/86 & 86/86\\
\hline
\hline
\hline
\#Validation Drug Pairs & 18,841 & 18,383 & 17,884\\
\hline
\#Unique Labels in Val & 86/86 & 86/86 & 86/86\\
\hline
\hline
\hline
\#Testing Drug Pairs & 3322 & 7899 & 12,893\\
\hline
[Test Pairs] 1 Drug not seen & 3312 & 7824 & 12,671\\
\hline
[Test Pairs] 2 Drugs not seen & 10 & 75 & 222\\
\hline
\#Unique Labels in Testing & 39/86 & 66/86 & 67/86\\
\hline
\end{tabular}
\caption{Drug Held-off partitioning statistics.}
\label{tab:heldoffparttable}
\end{table}

\begin{table*}[!htb]
\centering
\small
\resizebox{\textwidth}{!}{\begin{tabular}{|p{2.5cm}|p{1.5cm}|p{1.5cm}|p{1.5cm}|p{1.5cm}|p{1.5cm}|p{1.5cm}|p{1.5cm}|p{1.5cm}|}
\cline{1-9}
\textbf{Column} & \multicolumn{2}{|c|}{\bf{DDI Partition}} &  \multicolumn{2}{|c|}{\bf{1\% Drug Heldoff}} &  \multicolumn{2}{|c|}{\bf{2\% Drug Heldoff}} &  \multicolumn{2}{|c|}{\bf{3\% Drug Heldoff}}\\
\cline{2-9}
& Count = 1 & Count = 2 & Count = 1 & Count = 2 & Count = 1 & Count = 2 & Count = 1 & Count = 2\\
\cline{1-9}
ATC Code & 132 & 123 & 132 & 123 & 132 & 121 & 132 & 122\\
\hline
Categories & 2262 & 1659 & 2256 & 1646 & 2253 & 1653 & 2241 & 1645\\
\hline
Description & 12179 & 5905 & 12138 & 5873 & 12094 & 5862 & 12016 & 5800\\
\hline
Merged Class. & 710 & 565 & 712 & 566 & 710 & 561 & 698 & 558\\
\hline
Protein Binding & 1471 & 477 & 1467 & 479 & 1464 & 476 & 1452 & 472\\
\hline
Target Action & 3157 & 1539 & 3151 & 1538 & 3145 & 1488 & 3143 & 1524\\
\hline
\end{tabular}}
\caption{Vocabulary size (including 3 fixed tokens). \looseness=-1}
\label{tab:vocabsize}
\end{table*}

In our setup, we vary the percentage of drugs held-off i.e., `x' as 1\%, 2\% and 3\% to obtain different partitions, and report some of the key characteristics of the resulting data in Table~\ref{tab:heldoffparttable}.
Clearly, as x\% increases, the number of held-off test drugs increase, and so does the size of the final testing drug pairs.
The distribution of DDI pairs as well as the size of the training/validation set is different from the previous setting (described in Section~\ref{subsubsection:ddipartition}), and hence none of the results across these two separate partitioning schemes are comparable.
Also, note that, the label distribution of the test set is very different from that of the training/validation set, due to the way the splits have been constructed.
This skew in label distribution makes classifier performance comparison very difficult between the validation and testing sets within each of the x\% settings.
Hence, this partitioning scheme makes model training and evaluation tasks much more challenging, while also capturing a more realistic end-user scenario.\looseness=-1

\subsection{Classifier Baselines}
\label{subsection:baselines}

We compare the results from the LSTM based approaches against the bag-of-words baselines for the classification task.
To construct these baselines, the first step is to extract a bag-of-words representation for each column of each drug. 
First, we use the same vocabulary as the LSTM models for each column.
Now, each column can be represented in terms of the count of different tokens (from the vocabulary) in that column. 
We use the \texttt{scikit-learn}~\cite{scikit-learn} \texttt{CountVectorizer} for this purpose. 
Now, for each interaction pair, the pair can be represented as the sum of the count vectors of the corresponding columns for each drug.
This representation for a drug pair can be used as an input to multiple machine learning models. 
We run experiments using the popular k-nearest neighbors (KNN) and random forests (RF) models.
One notable difference between these count based representations and the ones derived from the LSTM is that LSTM encoding vectors are semantic vectors, which can be utilized for running Analogy SQL queries on databases.
The bag-of-word based approaches described here do not generate any encodings.
Thus the comparison of the BOW models against LSTM is only possible at the individual column level and that too for only the DDI prediction task, thereby greatly limiting their usage.\looseness=-1

Besides the above bag-of-words models, we additionally implement a random classifier as a baseline to demonstrate how much ``learning'' is truly happening for each of our models in presence of the skewness in DDI label distribution. 
Our ``Random'' classifier first computes the frequency distribution of each class label using the training data.
Then for each drug pair in validation/testing set, it randomly samples a class label using the frequency distribution computed above and compares the sampled label against the ground truth label for that pair to compute the accuracy.
Note that this classifier does not use any column specific information, and hence the performance is same for all columns on a particular data.
For each scenario, we run multiple simulations of the ``Random'' classifier to report the mean of the results.\looseness=-1

\subsection{Optimal Hyper-parameter Search}
\label{subsection:gridsearch}

We conduct a hyper-parameter search for each column independently on the BOW baselines as well as the LSTM model. 
For the BOW baselines, we use 3-fold cross validation on the concatenated training and validation data to select the best hyper-parameter.
We tune the value of $K$ for the KNN model, while the value of the number of estimators, max tree depth and  class weighted v/s non weighted loss are varied for the RF model.
For the LSTM model, cross validation is too expensive and hence each instance of the model is trained on the Training set and the best hyper-parameters are selected based on the performance on the Validation set using Accuracy as the scoring metric.
During hyper-parameter search, we vary the token embedding dimension, number of layers, hidden units within the LSTM cell, and the number of tokens considered.\looseness=-1

\subsection{Simulating Analogy Query}
\label{subsection:analogysimulation}

The best way to verify the efficacy of the Analogy SQL query~\cite{Bordawekar2017} for the drug database setting is a controlled user study involving a group of subject matter experts.
However, such a study is very costly, time-consuming and hard to setup due to the sheer lack of experts.
Thus, in the absence of an user study, we attempt to design an automated simulation and evaluation approach for Analogy SQL under supervised setting.
The main Analogy SQL query we study is : \newline
[A : B :: A : (D?)] Given the drugs A \& B, find other drugs (D) in the database that interact with drug A in a similar way as the drug A interacts with drug B.\newline
We lever the labelled drug pairs, already partitioned into validation and testing sets, to simulate such queries using Algorithm~\ref{alg:firstanalogy}, and evaluate the qualitative performance using the strategy described in Section~\ref{subsubsection:analogyevaluation}.\looseness=-1

\begin{algorithm}[!htb]
\caption{Simulation of [A : B :: A : (D?)]}
\label{alg:firstanalogy}  
\begin{algorithmic}[1]          
\REQUIRE Fully initialized Database
\REQUIRE Input DDI pairs of Val/Test partition
\FOR {(D1, D2) with label L in VAL/TEST DDI pairs} 
      \STATE C = count how many times D1 interacts with another drug (except D2) having interaction type L
      \IF{C == 0}
      \STATE Skip simulation with (D1, D2)
      \ELSE
      \STATE Use (D1, D2) to frame the analogy [D1 : D2 :: D1 : (D3?)] using Analogy SQL (Section~\ref{subsection:leveragingencoding})\looseness=-1
      \STATE Execute query on the DrugBankFullClean table using specific column encodings and fetch the list of D3 as D3List
      \STATE Compute the count of correct drugs (M) in D3List:\newline
    	SELECT COUNT(*) FROM DDITypeInfoTable\newline
    	    WHERE (LABEL = L) AND\newline
    	    (drug1=`D1' AND drug2 in D3List);
      \STATE  Precision@K = M/K
    \ENDIF
\ENDFOR
\RETURN
\end{algorithmic}
\end{algorithm}

\subsection{Evaluation Approach}
\label{subsection:evalmetrics}

\subsubsection{Classifier Performance Evaluation}
\label{subsubsection:classifierevaluation}

Our classification task is a one-out-of-86 class classification problem for each drug pair, where we have ground-truth DDI labels~\cite{Ryu2018deep}.
The qualitative performance of the classifiers is evaluated by computing the Accuracy (A), Macro F1 (M) and Weighted F1 (W) metrics on the Validation and Testing sets of the respective data partitioning schemes (Section~\ref{subsection:datapartitioning}), using the scikit-learn.metrics\footnote{\url{https://scikit-learn.org/stable/modules/classes.html\#module-sklearn.metrics}} package. The metrics are defined as follows:
\begin{align*}
    A &= \frac{\text{Correct Predictions}}{\text{Total Predictions}}\\
    M &= \sum\limits_{c=1}^{C}\frac{F1_{c}}{C} \\
    W &= \sum\limits_{c=1}^{C}\frac{\left(\frac{N_c}{N}\right)  F1_{c}}{C}
\end{align*}
where $F1_c$ denotes the $F1$ score for class $c$, with $C$ classes overall, and $N_c$ denotes the number of  samples in class $c$, and $N = \sum_{c=1}^C N_c$. The Precision (P), recall (R) and F1 score for a single class is defined as:
\begin{align*}
    P_c &= \frac{TP_c}{TP_c + FP_c}\\
    R_c &= \frac{TP_c}{TP_c + FN_c}\\
    F1_c &= \frac{2 * (P_c * R_c)}{(P_c + R_c)}
\end{align*}
where TP, FP, FN denote True Positives, False Positives and False Negatives respectively.

\subsubsection{Analogy SQL Performance Evaluation}
\label{subsubsection:analogyevaluation}

The ground truth DDI label for all drug pairs are stored in the DDITypeInfoTable during the database initialization step (Section~\ref{subsubsection:dbinit}), with the following column names : (drug1, drug2, label).
Our understanding of the data shared by Ryu et al.~\cite{Ryu2018deep} prompts us to assume that the order of the drug in the drug pair is important and should be retained during the simulation and evaluation steps.
Thus the triple (A, B, L) for the drug pair (A,B) with the DDI label L does not necessarily imply that the triple (B, A, L) is valid, which leads us to design a more conservative evaluation strategy.
The candidate list of answer drugs (D3List) is generated using only the column encodings of drugs as described in Step-6 \& 7 in Algorithm~\ref{alg:firstanalogy}.
The SQL query in Step-8 of Algorithm~\ref{alg:firstanalogy} is used to compute how many of the drugs, obtained as part of Analogy SQL query in Step-7, interact with the query drug D1 in the same way as D1 interacts with D2 (i.e., have the same interaction label L).
The order of the individual drugs are maintained by assigning them to appropriate column names (drug1 v/s drug2) in the SQL query of Step-8 in Algorithm~\ref{alg:firstanalogy}.
Additionally, it is possible that our Analogy SQL query may retrieve some answer drugs in D3List, which interact with D1 via a label L$^\prime$, such that the labels L and L$^\prime$ maybe closely related in terms of the underlying biological condition.
However, if such a pair is absent in the ground truth DDI pair, we consider that pair as an incorrect result during Precision computation, due to our lack of subject matter expertise to systematically judge the relatedness of the labels L and L$^\prime$.

\section{Results and Analysis}
\label{section:results}

We have outlined two possible data partitioning approaches in Section~\ref{subsection:datapartitioning} that we utilize to demonstrate the efficacy of our approach.
For each partitioning scheme, we present the results for the classification as well as the Analogy SQL task.
Note that the BOW classifier models are used to primarily evaluate how good our DNN model learned to predict the DDI relationship between drug pairs. 
Since our data has significant skew in the label distribution, we use Accuracy (A), Macro F1 (M) and Weighted F1 (W) to compare the qualitative performance of the classifiers.
We assume that higher the classification accuracy of the proposed Bi-LSTM model, better is the quality of the DDI task-specific embedding generated by the model, which should also get reflected in the performance of the Analogy SQL queries through improved Precision@K value.
We train and evaluate all the models using the same data partitions, while varying minimum word occurrence count value between 1 and 2.
Note that a minimum occurrence count value of 1 indicates that no word is pruned i.e., all unique tokens have been retained in the final vocabulary.
Table~\ref{tab:vocabsize} lists the effective size of the vocabulary for these different count threshold values for the respective data partitioning approaches.

\subsection{Scenario A :  DDI Pair Partitioning}
\label{subsection:ddipairresults}

In this section, we discuss the empirical results for DDI pair partitioning strategy described in Section~\ref{subsubsection:ddipartition}.
This data partitioning strategy follows the general approach of evaluating classifiers in machine learning domain, where the training, validation and testing sets have similar distribution of class labels, thereby making classifier performance comparisons much more straightforward.

The performance of the DNN model as well as the other baseline models are presented in Table~\ref{tab:ddipartition_classifier_minwc_1} and Table~\ref{tab:ddipartition_classifier_minwc_2} for the minimum word occurrence count values of 1 and 2 respectively.
The Random label selection algorithm does not depend on the text content of the corresponding column and hence the performance metrics are same across of all columns for the validation and testing set respectively.
The gap in performance between the Random label selection strategy and every other learning based strategy clearly demonstrates the performance benefits of using supervised learning approach for the DDI prediction task.
We observe that the DNN model consistently outperforms all the baseline models for each of the columns on both the validation and testing set.
The DNN model has approximately 6x improvement in accuracy for columns like ATC codes, categories and description when compared to the random label selection algorithm.
After DNN model, the best performing classifier is the BOW based Random Forest (RF) model.
The RF model outperforms the KNN model, but RF consistently under-performs for all columns when compared to the DNN model on the validation and testing set.
However, RF is very competitive for the categories column, where it is about 7\% poorer in terms of accuracy on the testing set compared to the DNN model for the minimum word count values of 1 and 2.
The DNN model performs better than the RF model by atleast ~7\% and atmost ~14-17\% on the testing data across of columns for different word count values.
Additionally, we observe that the performance gap of the DNN model on the testing set for different minimum word occurrence count values of 1 and 2 is very small across of the columns, although the size of the vocabulary is very different for those different minimum count values.
This may be due to the fact that the common words, which occur more than once and hence part of the vocabulary, influence the DDI prediction performance much more than the rare words that get pruned for minimum count value of 2.
Note that the BOW models do not generate any ``semantic encoding vector'' as the input to the models is a simple count vector of the words and hence can be used only for the classification task.
The DNN model performs better than the best performing RF based BOW model for each of the respective columns, which implies that the column encodings generated by our DNN model is able to capture, atleast to some extent, meaningful information about the drug-drug interaction relationship, and may be utilized for the Analogy SQL task.\looseness=-1

\begin{figure*}[!htb]
    \centering
    \begin{subfigure}[tbp]{0.5\textwidth}
        \includegraphics[width=\textwidth]{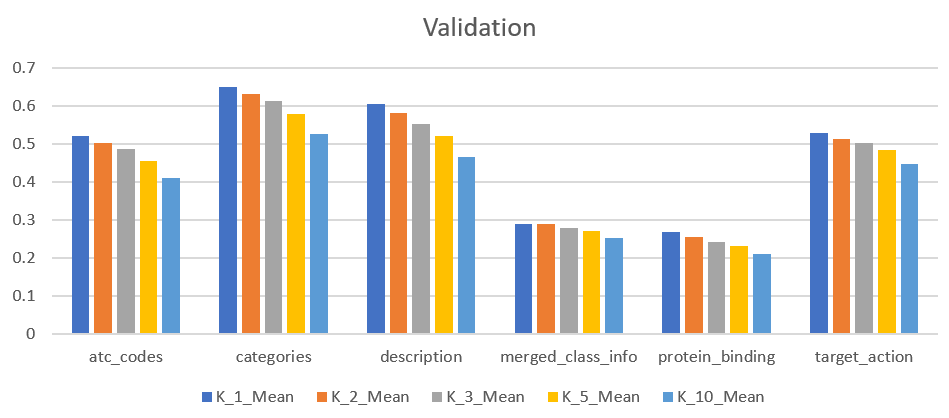}
        \label{fig:1centvalidation}
    \end{subfigure}%
    \begin{subfigure}[tbp]{0.5\textwidth}
        \includegraphics[width=\textwidth]{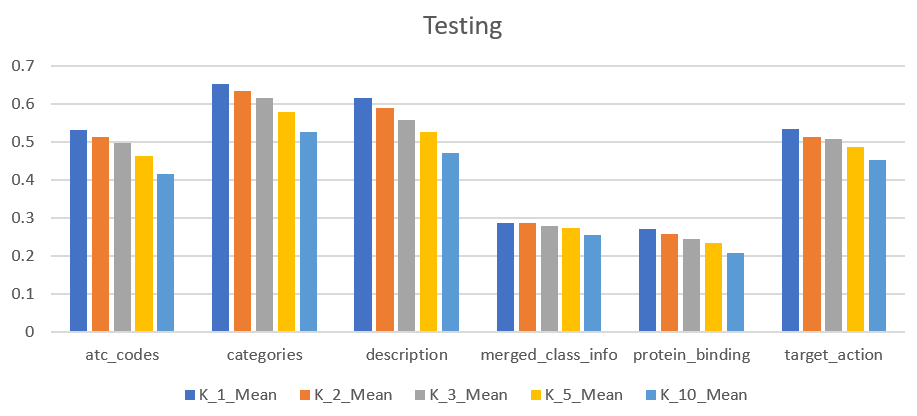}
        \label{fig:1centtesting}
    \end{subfigure}
    \caption{Precision@K for Analogy queries on respective data for DDI pair partitioning (Section~\ref{subsubsection:ddipartition}) for min-word-count = 1.\looseness=-1}
    \label{fig:ddipartition_analogy_type_1minwc_1}
\end{figure*}

\begin{figure*}[!htb]
    \centering
    \begin{subfigure}[tbp]{0.5\textwidth}
        \includegraphics[width=\textwidth]{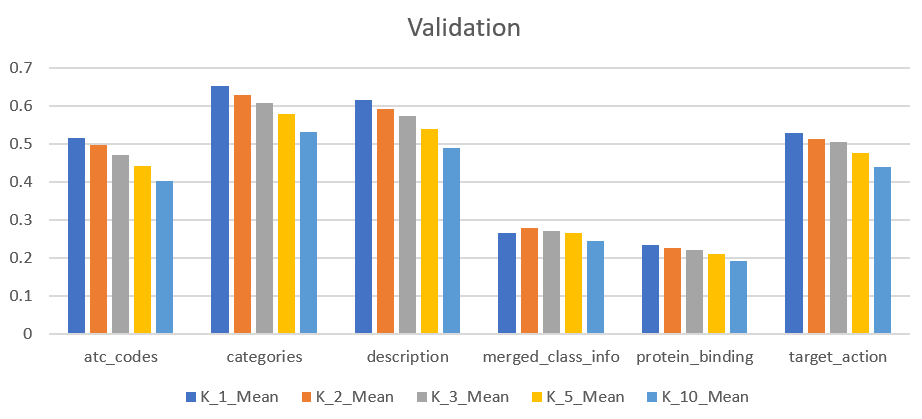}
        \label{fig:1centvalidation}
    \end{subfigure}%
    \begin{subfigure}[tbp]{0.5\textwidth}
        \includegraphics[width=\textwidth]{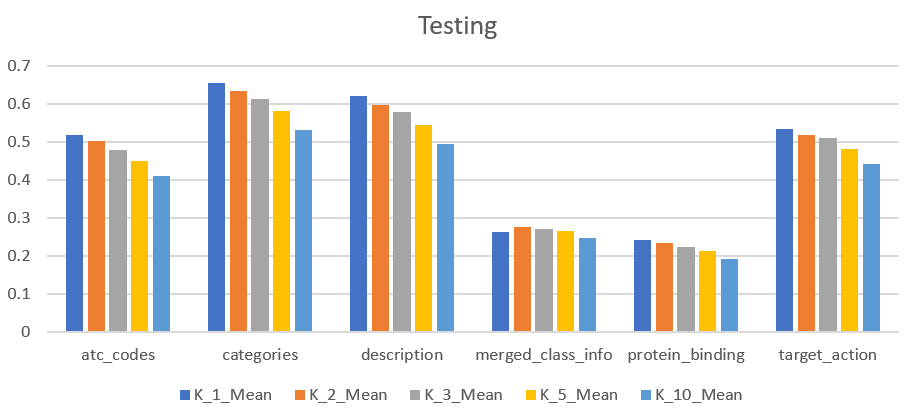}
        \label{fig:1centtesting}
    \end{subfigure}
    \caption{Precision@K for Analogy queries on respective data for DDI pair partitioning (Section~\ref{subsubsection:ddipartition}) for min-word-count = 2.\looseness=-1}
    \label{fig:ddipartition_analogy_type_1minwc_2}
\end{figure*}

We then simulate Analogy SQL on the validation and testing sets using Algorithm~\ref{alg:firstanalogy} by using the respective column encodings generated by the trained DNN model.
We vary the ``K'' to ${1,2,3,5,10}$ while computing the Precision@K for the Analogy SQL queries using the methodology defined in Section~\ref{subsubsection:analogyevaluation} and report the mean value obtained after simulating all the queries using the respective partitions.
Figure~\ref{fig:ddipartition_analogy_type_1minwc_1} and Figure~\ref{fig:ddipartition_analogy_type_1minwc_2} shows the performance of the Analogy SQL queries on the corresponding data using column encodings obtained by varying minimum word occurrence count value to 1 and 2 respectively.
For each column, the Precision@K is highest for K = 1, but then gradually decrease as K is increased to 10.
This means that the Analogy SQL queries are quite often able to fetch the top-most of the analogous drugs for the query pair correctly using our approach described in Algorithm~\ref{alg:firstanalogy}.
However, the Precision@K value generally decreases as K is gradually increased,implying that we do not consistently perform well for all top-K scenarios.
It is important to point out that some drugs may not have a total of K interactions of type L (query interaction label), which can contribute to this decrease in performance.
Due to the lack of an user study by subject matter experts, it is hard to determine how many of those ``incorrect'' answer drugs could be approximately ``semantically'' related in terms of the interaction label.
We observe that the columns, like ATC codes, categories and description, for which the DNN model achieves very high DDI prediction Accuracy value, show consistently higher performance in terms of Precision@K for the Analogy SQL simulations on both validation and testing set.
In contrast, some of the poorer performing columns, in terms of classification accuracy, like merged class information and protein binding, show very low Precision@K value for Analogy SQL simulations.
Thus, our original assumption that higher classification accuracy of the DNN model would lead to better quality of DDI task-specific embedding resulting in better performance of the Analogy SQL task, appears to hold in this case.\looseness=-1

\begin{table*}[!htb]
\centering
\large
\resizebox{\textwidth}{!}{\begin{tabular}{|p{2cm}|p{2cm}|p{2cm}|p{2cm}|p{2cm}|p{2cm}|p{2cm}|p{2cm}|p{2cm}|p{2cm}|p{2cm}|p{2cm}|p{2cm}|}
\cline{1-13}
Model & \multicolumn{2}{|c|}{\bf{ATC Codes}} & \multicolumn{2}{|c|}{\bf{Categories}}  & \multicolumn{2}{|c|}{\bf{Description}}  & \multicolumn{2}{|c|}{\bf{Merged Class}}  & \multicolumn{2}{|c|}{\bf{Protein Binding}}  & \multicolumn{2}{|c|}{\bf{Target Action}}\\
\cline{2-13}
& Validation & Testing & Validation & Testing & Validation & Testing & Validation & Testing & Validation & Testing & Validation & Testing\\
\cline{1-13}
Random & \textbf{A : 15.81} \newline M : 0.0117 \newline W : 0.1581 & \textbf{A : 15.70} \newline M : 0.0115 \newline W : 0.1569 & \textbf{A : 15.81} \newline M : 0.0117 \newline W : 0.1581 & \textbf{A : 15.70} \newline M : 0.0115 \newline W : 0.1569 & \textbf{A : 15.81} \newline M : 0.0117 \newline W : 0.1581 & \textbf{A : 15.70} \newline M : 0.0115 \newline W : 0.1569 & \textbf{A : 15.81} \newline M : 0.0117 \newline W : 0.1581 & \textbf{A : 15.70} \newline M : 0.0115 \newline W : 0.1569 &\textbf{A : 15.81} \newline M : 0.0117 \newline W : 0.1581 & \textbf{A : 15.70} \newline M : 0.0115 \newline W : 0.1569 & \textbf{A : 15.81} \newline M : 0.0117 \newline W : 0.1581 & \textbf{A : 15.70} \newline M : 0.0115 \newline W : 0.1569\\
\hline 
KNN & \textbf{A : 64.08} \newline M : 0.4391 \newline W : 0.6285 & \textbf{A : 64.20} \newline M : 0.4119 \newline W : 0.6290 & \textbf{A : 79.92} \newline M : 0.6601 \newline W : 0.7956 & \textbf{A : 80.17} \newline M : 0.7264 \newline W : 0.7980 & \textbf{A : 58.76} \newline M : 0.3418 \newline W : 0.5647 & \textbf{A : 58.61} \newline M : 0.3350 \newline W : 0.5630 & \textbf{A : 52.99} \newline M : 0.3273 \newline W : 0.5083 & \textbf{A : 52.29} \newline M : 0.3379 \newline W : 0.5016 & \textbf{A : 44.58} \newline M : 0.1437 \newline W : 0.4260 & \textbf{A : 43.99} \newline M : 0.1324 \newline W : 0.4194 & \textbf{A : 65.16} \newline M : 0.4914 \newline W : 0.6437 & \textbf{A : 65.29} \newline M : 0.4846 \newline W : 0.6444\\
\hline 
RF & \textbf{A : 76.80} \newline M : 0.5913 \newline W : 0.7619 & \textbf{A : 76.79} \newline M : 0.5772 \newline W : 0.7611 & \textbf{A : 89.85} \newline M : 0.7893 \newline W : 0.8964 & \textbf{A : 89.95} \newline M : 0.8258 \newline W : 0.8975 & \textbf{A : 79.67} \newline M : 0.5837 \newline W : 0.7898 & \textbf{A : 79.33} \newline M : 0.5973 \newline W : 0.7864 & \textbf{A : 60.47} \newline M : 0.4259  \newline W : 0.5926 & \textbf{A : 60.22} \newline M : 0.4358 \newline W : 0.5904 & \textbf{A : 50.56} \newline M : 0.1536 \newline W : 0.4625 & \textbf{A : 50.37} \newline M : 0.1656 \newline W : 0.4586 & \textbf{A : 74.50} \newline M : 0.5901 \newline W : 0.7388 & \textbf{A : 74.40} \newline M : 0.6125 \newline W : 0.7373\\
\hline 
DNN & \textbf{A : 92.15} \newline M : 0.7828 \newline W : 0.9203 & \textbf{A : 91.54} \newline M : 0.8051 \newline W : 0.9142 & \textbf{A : 96.88} \newline M : 0.8870 \newline W : 0.9686 & \textbf{A : 97.08} \newline M : 0.9332 \newline W : 0.9707 & \textbf{A : 96.35} \newline M : 0.8978 \newline W : 0.9633 & \textbf{A : 96.52} \newline M : 0.9367 \newline W : 0.9650 & \textbf{A : 72.79} \newline M : 0.5925 \newline W : 0.7223 & \textbf{A : 72.09} \newline M : 0.6035 \newline W : 0.7154 & \textbf{A : 61.66} \newline M : 0.3641 \newline W : 0.5972 & \textbf{A : 61.26} \newline M : 0.3841 \newline W : 0.5926 & \textbf{A : 89.42} \newline M : 0.7822 \newline W : 0.8931 & \textbf{A : 88.99} \newline M : 0.8027 \newline W : 0.8887\\
\hline 
\end{tabular}}
\caption{Qualitative performance comparison of DDI prediction by DNN and other baseline models on Validation and Testing set for DDI Pair partitioning (Section~\ref{subsubsection:ddipartition}) using min-word-count = 1.\looseness=-1}
\label{tab:ddipartition_classifier_minwc_1}
\end{table*}

\begin{table*}[!htb]
\centering
\large
\resizebox{\textwidth}{!}{\begin{tabular}{|p{2cm}|p{2cm}|p{2cm}|p{2cm}|p{2cm}|p{2cm}|p{2cm}|p{2cm}|p{2cm}|p{2cm}|p{2cm}|p{2cm}|p{2cm}|}
\cline{1-13}
Model & \multicolumn{2}{|c|}{\bf{ATC Codes}} & \multicolumn{2}{|c|}{\bf{Categories}}  & \multicolumn{2}{|c|}{\bf{Description}}  & \multicolumn{2}{|c|}{\bf{Merged Class}}  & \multicolumn{2}{|c|}{\bf{Protein Binding}}  & \multicolumn{2}{|c|}{\bf{Target Action}}\\
\cline{2-13}
& Validation & Testing & Validation & Testing & Validation & Testing & Validation & Testing & Validation & Testing & Validation & Testing\\
\cline{1-13}
Random & \textbf{A : 15.81} \newline M : 0.0117 \newline W : 0.1581 & \textbf{A : 15.70} \newline M : 0.0115 \newline W : 0.1569 & \textbf{A : 15.81} \newline M : 0.0117 \newline W : 0.1581 & \textbf{A : 15.70} \newline M : 0.0115 \newline W : 0.1569 & \textbf{A : 15.81} \newline M : 0.0117 \newline W : 0.1581 & \textbf{A : 15.70} \newline M : 0.0115 \newline W : 0.1569 & \textbf{A : 15.81} \newline M : 0.0117 \newline W : 0.1581 & \textbf{A : 15.70} \newline M : 0.0115 \newline W : 0.1569 &\textbf{A : 15.81} \newline M : 0.0117 \newline W : 0.1581 & \textbf{A : 15.70} \newline M : 0.0115 \newline W : 0.1569 & \textbf{A : 15.81} \newline M : 0.0117 \newline W : 0.1581 & \textbf{A : 15.70} \newline M : 0.0115 \newline W : 0.1569\\
\hline 
KNN & \textbf{A : 64.27} \newline M : 0.4369 \newline W : 0.6299 & \textbf{A : 64.36} \newline M : 0.4151 \newline W : 0.6299 & \textbf{A : 79.78} \newline M : 0.6569 \newline W : 0.7942 & \textbf{A : 80.14} \newline M : 0.7124 \newline W : 0.7976 & \textbf{A : 58.55} \newline M : 0.3424 \newline W : 0.5628 & \textbf{A : 58.48} \newline M : 0.3517 \newline W : 0.5624 & \textbf{A : 52.31} \newline M : 0.3277 \newline W : 0.5037 & \textbf{A : 52.03} \newline M : 0.3312 \newline W : 0.5009 & \textbf{A : 41.99} \newline M : 0.0638 \newline W : 0.3666 & \textbf{A : 41.91} \newline M : 0.0665 \newline W : 0.3684 & \textbf{A : 65.60} \newline M : 0.4901 \newline W : 0.6484 & \textbf{A : 65.29} \newline M : 0.4876 \newline W : 0.6447\\
\hline 
RF & \textbf{A : 76.80} \newline M : 0.5921 \newline W : 0.7620 & \textbf{A : 76.50} \newline M : 0.5742 \newline W : 0.7581 & \textbf{A : 89.97} \newline M : 0.7924 \newline W : 0.8977 & \textbf{A : 90.10} \newline M : 0.8366 \newline W : 0.8993 & \textbf{A : 80.09} \newline M : 0.6056 \newline W : 0.7947 & \textbf{A : 79.87} \newline M : 0.6196 \newline W : 0.7924 & \textbf{A : 59.77} \newline M : 0.4268 \newline W : 0.5865 & \textbf{A : 59.39} \newline M : 0.4284 \newline W : 0.5826 & \textbf{A : 49.03} \newline M : 0.1486 \newline W : 0.4459 & \textbf{A : 48.89} \newline M : 0.1526 \newline W : 0.4424 & \textbf{A : 74.32} \newline M : 0.5883 \newline W : 0.7370 & \textbf{A : 74.18} \newline M : 0.6173 \newline W : 0.7354\\
\hline 
DNN & \textbf{A : 91.89} \newline M : 0.8069 \newline W : 0.9176 & \textbf{A : 91.75} \newline M : 0.8184 \newline W : 0.9160 & \textbf{A : 96.89} \newline M : 0.9111 \newline W : 0.9686 & \textbf{A : 96.85} \newline M : 0.9328 \newline W : 0.9683 & \textbf{A : 96.68} \newline M : 0.9102 \newline W : 0.9667 & \textbf{A : 96.67} \newline M : 0.9300 \newline W : 0.9665 & \textbf{A : 72.23} \newline M : 0.5986 \newline W : 0.7187 & \textbf{A : 71.98} \newline M : 0.6115 \newline W : 0.7162 & \textbf{A : 61.57} \newline M : 0.3647 \newline W : 0.5964 & \textbf{A : 60.81} \newline M : 0.3739 \newline W : 0.5869 & \textbf{A : 89.15} \newline M : 0.8062 \newline W : 0.8906 & \textbf{A : 88.81} \newline M : 0.8074 \newline W : 0.8868\\
\hline 
\end{tabular}}
\caption{Qualitative performance comparison of DDI prediction by DNN and other baseline models on Validation and Testing set for DDI Pair partitioning (Section~\ref{subsubsection:ddipartition}) with min-word-count = 2.\looseness=-1}
\label{tab:ddipartition_classifier_minwc_2}
\end{table*}

\subsection{Scenario B : Drug Held-off Partitioning}
\label{subsection:drugheldoffresults}

In this section, we present the simulation results for a fixed percentage of the Drug held-off partitioning strategy described in Section~\ref{subsubsection:heldoffpartition}.
This is a more realistic end user scenario, where we train all our models using the ground truth label information of existing drugs, and then use such trained models to predict DDI with newly added drugs~\cite{Zhang2015label}.
However, evaluating this setup is very tricky as the class label distribution is no longer same across the training, validation and testing sets.
Under this setup, the training and validation sets have approximately the same class label distribution, but the testing set has a different label distribution, as observed in the Table~\ref{tab:heldoffparttable}.\looseness=-1

\begin{figure*}[!htb]
    \centering
    \begin{subfigure}[tbp]{0.5\textwidth}
        \includegraphics[width=\textwidth]{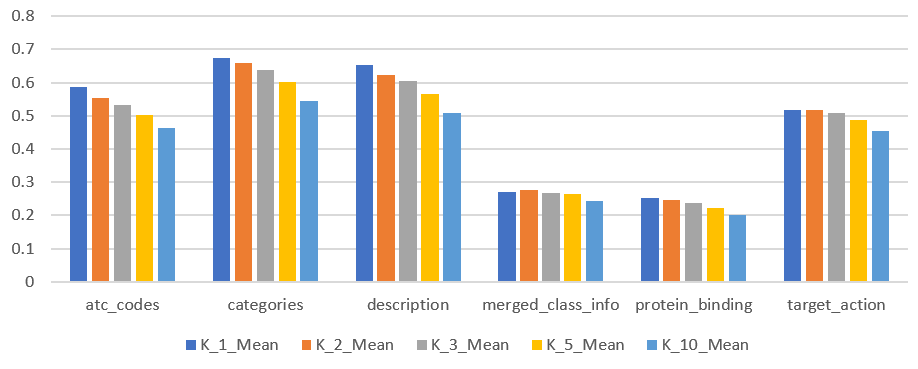}
        \caption{1\% Data - Validation}
        \label{fig:1centvalidation}
    \end{subfigure}%
    \begin{subfigure}[tbp]{0.5\textwidth}
        \includegraphics[width=\textwidth]{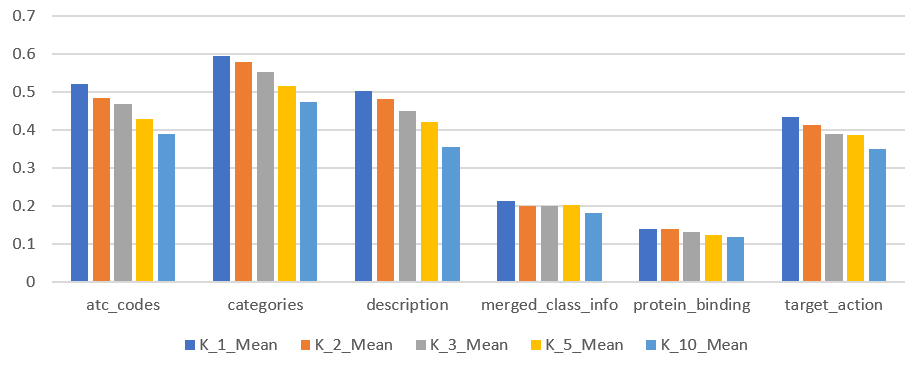}
        \caption{1\% Data - Testing}
        \label{fig:1centtesting}
    \end{subfigure}
    
    \begin{subfigure}[tbp]{0.5\textwidth}
        \includegraphics[width=\textwidth]{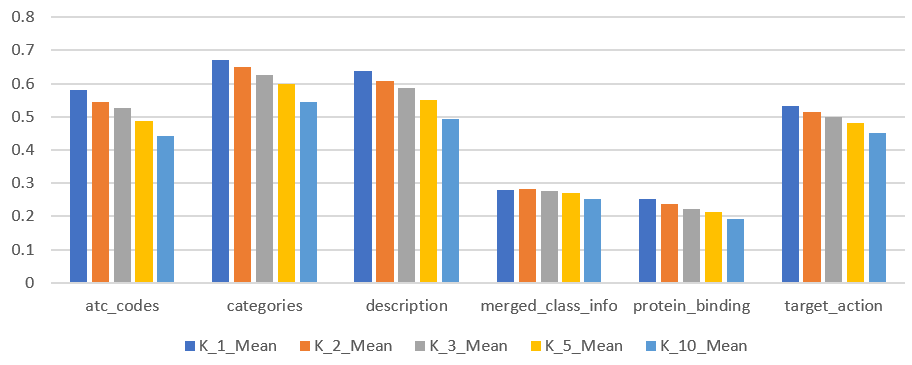}
        \caption{2\% Data - Validation}
        \label{fig:2centvalidation}
    \end{subfigure}%
    \begin{subfigure}[tbp]{0.5\textwidth}
        \includegraphics[width=\textwidth]{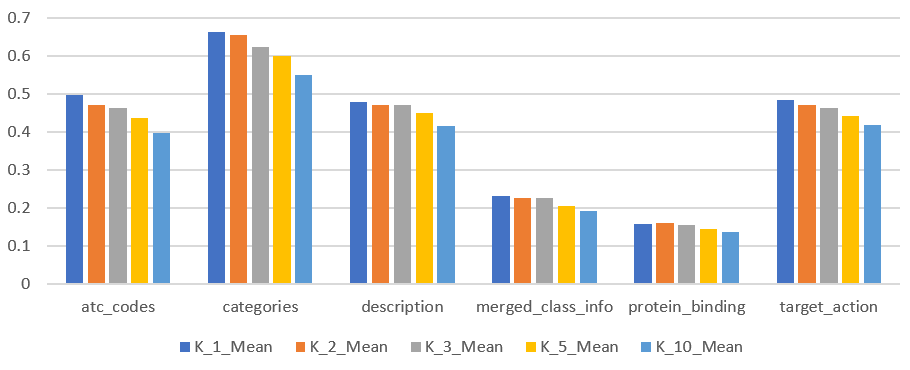}
        \caption{2\% Data - Testing}
        \label{fig:2centtesting}
    \end{subfigure}
    
    \begin{subfigure}[tbp]{0.5\textwidth}
        \includegraphics[width=\textwidth]{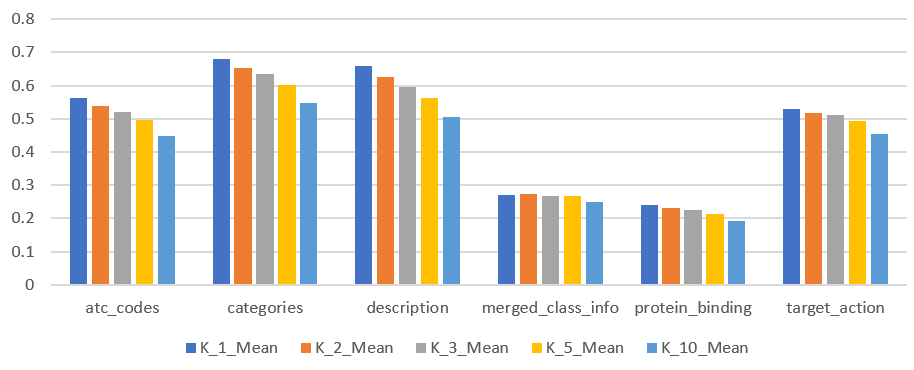}
        \caption{3\% Data - Validation}
        \label{fig:2centvalidation}
    \end{subfigure}%
    \begin{subfigure}[tbp]{0.5\textwidth}
        \includegraphics[width=\textwidth]{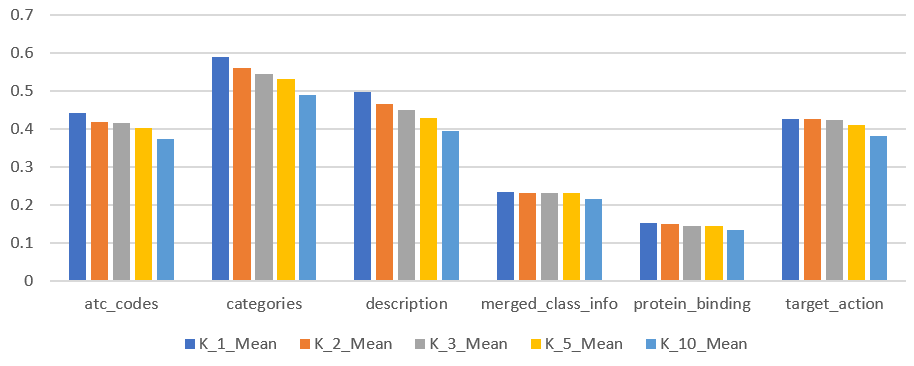}
        \caption{3\% Data - Testing}
        \label{fig:2centtesting}
    \end{subfigure}
    \caption{Precision@K for Analogy queries on respective data for different percentage of Drug Held-off partitioning (Section~\ref{subsubsection:heldoffpartition} for min-word-count = 2.}\label{fig:analogy_sql_drug_held_off_minwc_2}
\end{figure*}

For brevity, we present the classifier performance results for 2\% drug held off setting in Table~\ref{tab:ddipartition_2_cent_classifier_minwc_2}, and the performance of Analogy SQL queries on 1\%, 2\% and 3\% drug held partitioning in Figure~\ref{fig:analogy_sql_drug_held_off_minwc_2}, for a minimum word count value of 2.
In terms of testing accuracy, we observe that the Random Forest model performs better than the DNN model for the categories and the description columns, while the DNN model performs better for ATC codes, merged class and protein binding, whereas the performance on the target action column is almost same.
We have observed similar change in performance between the Random Forest and DNN model for some of the columns with other drug held-off percentage values as well.
This observation leads us to conclude that under different drug held-off settings, our proposed DNN model may not always be able to generate high quality  encodings for certain columns, as indicated by a lower Accuracy score than the Random Forest model.
Interestingly, the Random Forest model can still be used for achieving better DDI prediction accuracy, if that is the only requirement.
For different percentages of drug held-off partitioning, while the overall performance trends for the Analogy SQL query appear very similar, there is a significant dip in performance, specifically for columns like protein binding and merged class information.
Note that the DNN classifier performance on the testing set have also degraded for these columns compared to the performance on the validation set, which explains this relatively poor performance.
Also, the Precision@K performance in Validation set is slightly better (by 0.1) than that in the Testing set in some cases (notice the difference in y-axis value).
Nonetheless, our results on the 2\% data (testing set) are very encouraging and we believe this can lead to improved results with additional future investigations.

\begin{table*}[!htb]
\centering
\large
\resizebox{\textwidth}{!}{\begin{tabular}{|p{2cm}|p{2cm}|p{2cm}|p{2cm}|p{2cm}|p{2cm}|p{2cm}|p{2cm}|p{2cm}|p{2cm}|p{2cm}|p{2cm}|p{2cm}|}
\cline{1-13}
Model & \multicolumn{2}{|c|}{\bf{ATC Codes}} & \multicolumn{2}{|c|}{\bf{Categories}}  & \multicolumn{2}{|c|}{\bf{Description}}  & \multicolumn{2}{|c|}{\bf{Merged Class}}  & \multicolumn{2}{|c|}{\bf{Protein Binding}}  & \multicolumn{2}{|c|}{\bf{Target Action}}\\
\cline{2-13}
& Validation & Testing & Validation & Testing & Validation & Testing & Validation & Testing & Validation & Testing & Validation & Testing\\
\cline{1-13}
Random & \textbf{A : 15.91} \newline M : 0.0117 \newline W : 0.1591 & \textbf{A : 14.37} \newline M : 0.0116 \newline W : 0.1380 & \textbf{A : 15.91} \newline M : 0.0117 \newline W : 0.1591 & \textbf{A : 14.37} \newline M : 0.0116 \newline W : 0.1380 & \textbf{A : 15.91} \newline M : 0.0117 \newline W : 0.1591 & \textbf{A : 14.37} \newline M : 0.0116 \newline W : 0.1380 & \textbf{A : 15.91} \newline M : 0.0117 \newline W : 0.1591 & \textbf{A : 14.37} \newline M : 0.0116 \newline W : 0.1380 & \textbf{A : 15.91} \newline M : 0.0117 \newline W : 0.1591 & \textbf{A : 14.37} \newline M : 0.0116 \newline W : 0.1380 & \textbf{A : 15.91} \newline M : 0.0117 \newline W : 0.1591 & \textbf{A : 14.37} \newline M : 0.0116 \newline W : 0.1380 \\
\hline 
KNN & \textbf{A : 66.95} \newline M : 0.4759 \newline W : 0.6620 & \textbf{A : 53.15} \newline M : 0.3628 \newline W : 0.5214 & \textbf{A : 80.18} \newline M : 0.7276 \newline W : 0.7980 & \textbf{A : 75.93} \newline M : 0.7071 \newline W : 0.7533 & \textbf{A : 58.89} \newline M : 0.3702 \newline W : 0.5663 & \textbf{A : 53.77} \newline M : 0.3425 \newline W : 0.5126 & \textbf{A : 54.95} \newline M : 0.4103 \newline W : 0.5420 & \textbf{A : 32.40} \newline M : 0.2362 \newline W : 0.3278 & \textbf{A : 49.15} \newline M : 0.1336 \newline W : 0.4425 & \textbf{A : 34.71} \newline M : 0.1106 \newline W : 0.2831 & \textbf{A : 67.19} \newline M : 0.5289 \newline W : 0.6630 & \textbf{A : 57.61} \newline M : 0.4293 \newline W : 0.5584\\
\hline 
RF & \textbf{A : 77.11} \newline M : 0.6123 \newline W : 0.7659 & \textbf{A : 62.37} \newline M : 0.5288 \newline W : 0.6028 & \textbf{A : 90.64} \newline M : 0.8527 \newline W : 0.9044 & \textbf{A : 81.85} \newline M : 0.7496 \newline W : 0.8111 & \textbf{A : 80.73} \newline M : 0.6301 \newline W : 0.8013 & \textbf{A : 68.73} \newline M : 0.5656 \newline W : 0.6613 & \textbf{A : 60.76} \newline M : 0.4461 \newline W :0.5950 & \textbf{A : 36.84} \newline M : 0.2672 \newline W : 0.3655 & \textbf{A : 49.36} \newline M : 0.1556 \newline W : 0.4465 & \textbf{A : 36.06} \newline M : 0.1276 \newline W : 0.2875 & \textbf{A : 75.10} \newline M : 0.6459 \newline W : 0.7448 & \textbf{A : 61.34} \newline M : 0.4985 \newline W : 0.5968\\
\hline 
DNN & \textbf{A : 91.92} \newline M : 0.8045 \newline W : 0.9176 & \textbf{A : 65.53} \newline M : 0.4901 \newline W : 0.6385 & \textbf{A : 96.22} \newline M : 0.9250 \newline W : 0.9620 & \textbf{A : 78.15} \newline M : 0.7083 \newline W : 0.7757 & \textbf{A : 96.17} \newline M : 0.9363 \newline W : 0.9616 & \textbf{A : 65.13} \newline M : 0.4814 \newline W : 0.6409 & \textbf{A : 73.43} \newline M : 0.6053 \newline W : 0.7294 & \textbf{A : 41.18} \newline M : 0.3077 \newline W : 0.4092 & \textbf{A : 62.30} \newline M : 0.3620 \newline W : 0.6009 & \textbf{A : 41.26} \newline M : 0.1582 \newline W : 0.3570 & \textbf{A : 89.38} \newline M : 0.8342 \newline W : 0.8926 & \textbf{A : 61.39} \newline M : 0.4579 \newline W : 0.5986\\
\hline 
\end{tabular}}
\caption{Qualitative performance comparison of DDI prediction by DNN and other baseline models on Validation and Testing set for 2\% Drug Held-off partitioning (Section~\ref{subsubsection:heldoffpartition}) with min-word-count = 2.\looseness=-1}
\label{tab:ddipartition_2_cent_classifier_minwc_2}
\end{table*}

\subsection{Discussions and Future Work}
\label{subsection:discussions}

Our approach is built to lever gold-standard label information to generate task-specific supervised database column encodings.
We observe that for some columns (like ATC codes and categories) our approach achieves very high qualitative performance on both the classification as well as the Analogy SQL task for the DDI Pair partitioning approach, while the performance is competitive for the Drug Held-off partitioning approach.
The performance drops drastically for some other columns (like merged class information and protein binding) across all settings.
This drop in performance could be due to either lack of adequate task-specific information in these columns or our tokenization and information encoding scheme being not effective for that specific column.
We think that column-specific textification and tokenization scheme, which can balance between human readability and sub-word based information reuse~\cite{Gage1994}, along with more advanced encoders (like LSTM with Attention~\cite{Zheng2017attention, Yang2016}) may improve the performance for some of these under-performing columns.

An obvious criticism of our approach would be that it is focused on utilizing a single column's information and hence cannot utilize information from across columns.
We argue that each column often has different data types as well as diverse information and may contain a different perspective on semantic similarity, which our approach will be able to capture better.
For example, the ATC codes column~\footnote{\url{https://en.wikipedia.org/wiki/Anatomical_Therapeutic_Chemical_Classification_System}} contains codes, each of which are of fixed length, and the codes capture an implicit clustering information of drugs, in terms of the organ/system of action, besides their therapeutic intent and other characteristics.
This column is interpretable only to an expert who is familiar with ATC coding system.
On the other hand, the description column contains human readable free text,making it much more interpretable to ordinary users, but may contain too many details that may not necessarily be relevant.
Thus our DNN model for the ATC codes column tries to capture the implicit clustering based interaction information, while the DNN model for the description column tries to capture the more explicit text based (i.e., medical concept based) interaction information.
Clearly, when the encodings of these two columns are separately used for Analogy SQL task, they provide very different perspectives for the drug drug interaction prediction task, which may be useful for better debugging during execution of end user scenarios.

While there is room for further improvement in DDI prediction and Analogy SQL performances for some of the columns, many other columns of the drug database have been currently unexplored.
Note that the drug database has a wide variety of data types like numeric and chemical structure information in its columns.
One approach could be to follow the textification and tokenization strategy proposed by Bordawekar et al. ~\cite{Bordawekar2017} for all columns like numeric, categorical and text based columns.
However, for a column like chemical structure that contains a graph representing the chemical structure of the drug (often stored in SMILES format~\footnote{\url{https://en.wikipedia.org/wiki/Simplified_molecular-input_line-entry_system}} in databases), it may be a better idea to encode the information through specialized models like graph convolutional network~\cite{Zitnik2018modeling} instead of using the textification approach~\cite{Bordawekar2017}.
This require data type specific encoders for individual columns (graph v/s text encoders) thereby making the models more complex as well as significantly increasing the model maintenance cost, rather than the current simplistic text encoder based strategy that follows the broad textification approach proposed by Bordawekar et al.~\cite{Bordawekar2017}.
However, we think that such data type specific encoders will be able to better capture the latent information of the specific data type, despite the increase in model cost and complexity, and hence needs future exploration.

The Analogy SQL task that we have studied is [A : B :: A : (D?)]. 
However, another practical and even more challenging analogy query would be [A : B :: C : (D?)], where we seek a drug D which interacts with drug C in the same way as A interacts with B.
Supervised evaluation of this scenario through simulation is quite challenging, as we need to have a systematic approach to generate the query triples (A, B, C) and ensure that we have a consistent definition of the correct answer drug D for the query triple (A, B, C).
In this scenario, it would be ideal to conduct an user study of our framework by recruiting subject matter experts and enabling them via a web portal to test their hypothesis using our column encoding based Analogy SQL queries.
Moreover, since our Analogy SQL queries are approximate by nature, there will always be some errors in the final result.
A portion of this error could be due to answer drugs being retrieved from a class label (L$^\prime$), which is semantically related to the query interaction label (L), but we currently flag those as incorrect results, due to the lack of a principled way to compute the label similarities.
An user study by subject matter experts can potentially reveal this error source in our approach, giving us more insight and possible scope of further improvement.
However, recruiting experts and performing such an user study is still challenging in terms of cost and time, that we could not address as part of this work.\looseness=-1

One shortcoming of our current approach is that we assume all the tables in the database are static while generating the corresponding column encodings.
However, database updates may be quite frequent~\cite{Bordawekar2017} and can be typically of two types in our settings, viz: 1) update in drug information (i.e., database column values) 2) update in drug drug interaction information (old interactions removed, new interactions added etc.).
As the information gets modified in different tables, the overall semantic context of the columns, captured by the supervised column encodings, should also get updated accordingly, similar to the updates proposed by Bordawekar et al.~\cite{Bordawekar2017} for the unsupervised model.
Unfortunately, modifying supervised encodings for database updates could be a non-trivial task.
In our supervised setting, intuitively, the newly added interaction of a drug pair may be easier to handle by incrementally re-training the DNN model first, using the new interactions alongside a subset of existing interactions for the corresponding drugs, and then by regenerating the column encodings for all drugs using the updated DNN model.
But deleted interaction between any drug pair is very difficult to incrementally ``unlearn'', and even harder to evaluate the deletion.
Similarly, any update to the column values for any drug will require re-training of our DNN model, using atleast each such interaction pairs where the drug with updated information appear.
Since we have one model per-column, updates to individual columns would mean re-training of the column-specific model, which may be less costly sometimes, provided the column data is small and the number of columns updated is small too.
All these are intuitive approaches which will need additional empirical validations.
Note that, this assumption of static information may be less problematic for our Drug database scenario, due to relatively low frequency of updates in practice, but will be a major issue for other medical databases like EHR data of patients in hospitals (e.g.: MIMIC III by \cite{mimiciii}), where all types of database updates are very frequent.
Thus, designing task-specific  supervised column encodings for databases with rapid and diverse updates is a very challenging problem, and is worth a thorough investigation in future. \looseness=-1
\section{Conclusion}
\label{section:conclusion}

We study the task of semantic information preserving supervised column encoding generation of multi-token text columns of a relational database.
We use the Drug Drug Interaction prediction scenario as a case study, whereby we seek to learn the column encodings of a pair of rows of the drug information table through supervised learning method, using the ground truth DDI label information for the drug pair.
We propose a DNN model for the text-based DDI prediction task, which has minimum text processing overhead compared to previous works, and our proposed model achieves a very high DDI prediction accuracy.
Additionally, we utilize those column encodings to simulate Analogy SQL query on the relational database and propose an evaluation strategy to demonstrate the efficacy of the column encodings for the Analogy SQL task.\looseness=-1

\balance
\bibliographystyle{abbrv}
\bibliography{paper}

\end{document}